\documentclass[12pt,dvips]{article}
\usepackage{amsmath,amssymb}
\usepackage[dvips]{graphicx}
\usepackage{hyperref}

\begin{document} 

\begin{titlepage}

\title{Monte Carlo Simulation of Laser Diodes Sub-Poissonian Light 
Generation}

\author{Laurent Chusseau\thanks{E-mail: chusseau@univ-montp2.fr},
Jacques Arnaud\thanks{Mas Liron, F30440 Saint Martial, France.} \\
{\normalsize Centre d'\'Electronique et de Micro-opto\'electronique de
Montpellier} \\
{\normalsize UMR 5507 CNRS, Universit\'e Montpellier II} \\
{\normalsize F34095 Montpellier, France.}}

\maketitle

\begin{abstract} 
When laser diodes are driven by high-impe\-dance electrical sources
the variance of the number of photo-detection events counted over
large time durations is less than the average number of events
(sub-Poissonian light).  The paper presents a Monte Carlo simulation
that keeps track of each level occupancy (0 or 1) in the conduction
and valence bands, and of the number of light quanta in the optical
cavity.  When there is good electron-lattice thermal contact the
electron and hole temperatures remain equal to that of the lattice. 
In that case, elementary laser-diode noise theory results are
accurately reproduced by the simulation.  But when the thermal
contact is poor (or, almost equivalently, at high power levels) new
effects occur (spectral-hole burning, temperature fluctuations,
statistical fluctuations of the optical gain) that are difficult
to handle theoretically.  Our numerical simulation shows that the
frequency domain over which the photo-current spectral density is
below the shot-noise level becomes narrower as the optical power 
increases.
\end{abstract}

\end{titlepage}

\section{Introduction}

Light is called sub-Poissonian when the variance of the number of
photo-detection events counted over a large time duration is less than
the average number of events.  Equivalently, we may say that the
photo-current spectral density is below the shot-noise level at low
Fourier (or baseband) frequencies.  It has been shown experimentally
in 1987 \cite{yam1} that laser diodes driven by high-resistance
electrical sources may generate sub-Poissonian light.  This feature of
great fundamental and practical importance treated theoretically in
\cite{yam1} on the basis of the laws of Quantum Optics may be
understood alternatively as resulting from a birth-death Markov
process \cite{arn1,arn2}.  Such processes, as they apply to lasers in
general, were discussed in particular by Jakeman and Loudon
\cite{jakeman, loudon}.  The reader will find a clear mathematical
discussion in \cite{gillespie}.

The originality of the present simulation rests on the fact that
electronic level occupancies and light quanta are accounted for
\emph{individually}, the photo-detection rate being derived from a
Monte Carlo simulation of the system evolution in the course of time. 
All microscopic physical processes are (or could be) accounted for. 
Electron level occupancies (0 or 1) fluctuate as a result of internal
electronic transitions (Auger transitions, thermalization and
interaction with the optical field stored in the laser cavity),
pumping and optical absorption.

An exhaustive Monte Carlo simulation has apparently not been reported
before, probably because of the high computer time required in the
case of bulk semiconductor lasers.  But dramatic size reductions have
been obtained with microcavity quantum dot lasers \cite{ando} and
two-dimensional photonic-band\-gap quantum-dots lasers \cite{seassal}. 
Because the active layers involve few optically active quantum dots in
the gain region an individual account of each level occupancy is
manageable.  Such devices, likely to be the next generation
semiconductor lasers provided technological problems are overcome to
achieve efficient electrical pumping, may be labeled `mesoscopic'
because of their small size.  Note that in optical computing
applications only small optical output powers are needed and small
driving currents are desired.

For conventional applications, e.g., optical communications,
milliwatts of powers are usually required.  In that case, our
simulation results have to be scaled up since it would be impractical
to account for every level occupancy in the case of bulk
semiconductors.  Scaling laws applicable to the linearized theory need
generalization if one wishes to take into account advanced effects
such as spectral-hole burning.  Our numerical results help ascertain
the validity of advanced theories.

The main processes involved in laser light generation are recalled in
Section \ref{process}, and the numerical procedure is explained in
Section \ref{birth-death}.  Numerical results concerning
photodetection noise are illustrated in Section \ref{results} and the
conclusion in Section \ref{conclusion}.  The analytical formula in the
appendix derives from the birth-death process through linearization
and the assumption that the Fermi-Dirac distribution at some constant
temperature $T$ adequately describes electronic level occupancies.  An
explicit expression of the photo-detection spectrum is obtained for
evenly-spaced electronic levels (spacing $\epsilon$) and a
non-fluctuating electron injection rate $J$.  It is shown that the
photo-detection spectrum depends only on the light absorption constant
$\alpha$ and the normalized pumping rate $J^{\star} \equiv J \epsilon
/ k_{B} T$.

\section{Laser Diode Processes}
\label{process}

In the present section the basic Physics of semiconductors, and
processes relevant to isolated cavities containing semiconductors in
contact with a thermal bath, are first recalled.  Pumping and optical
absorption are discussed subsequently.

One-electron energy levels in the semiconductor are supposed to be of
the form $\epsilon_{k} = k \epsilon $, with $k$ an integer and
$\epsilon$ a constant.  The lasers considered may incorporate quantum
dots in the gain region.  The evenly-spaced-level assumption is then
justified by the mechanism of level `repulsion' observed in
nanometer-scale irregular particles \cite{denton}: the probability
that adjacent levels be separated by $\epsilon$ is of the form
$\epsilon^4 \exp(-\epsilon^2)$, a sharply peaked function of
$\epsilon$.  Note also that quantum wells exhibit levels that are
evenly spaced \emph{on the average} within each sub-bands.  That is,
the density of states is a constant.

Some levels are allowed while others are forbidden.  Allowed levels
may be occupied by at most one electron to comply with the Pauli
exclusion principle.  The electron spin, ignored in the present paper
for the sake of brevity, is discussed for example in \cite{notreprb}. 
Our model considers neither electronic superposition states nor any
strong Coulomb interaction, approximations made in virtually all
laser-diode theories.  In semiconductors the allowed electronic levels
group into two bands, the upper one being called \emph{conduction}
band (CB) and the lower one \emph{valence} band (VB).  We suppose that
both bands involve the same number $B = 100$ of levels, and are
separated by $G_{p}$ forbidden levels as shown in Fig.~\ref{fig1}. 
The band gap energy $G_{p} \epsilon$ is instrumental in determining
the laser oscillation frequency but it will not enter in our model
because of simplifying assumptions to be later discussed.  Only one
valence band is considered, but it would be straightforward to take
into account the heavy-hole, light-hole and split-off bands found in
most semiconductors.  $N = B$ electrons are allocated to the allowed
energy levels.  For pure semiconductors at $T = 0$~K, the $N$ electrons
fill up the valence band while the conduction band is empty.  The
electron-lattice system is electrically neutral.

Without a thermal bath, the electron gas reaches an equilibrium state
through Auger-type transitions: an electron gets promoted to upper
levels while another electron gets demoted to lower levels in such a
way that the total energy remains the same.  The two electrons may
belong to the same band or to distinct bands, but only the former
situation is presently considered.  Auger transitions ensure that all
the system microstates are being explored in the course of time so
that electron gases possess well-defined temperatures at any instant
in each band.  Nothing, however, prevents these temperatures from
fluctuating in the course of time.  The Coulomb interactions on which
Auger processes rely is supposed to be weak, so that one-electron
level schemes are applicable.  We have simulated energy-conserving
processes and recovered recently-reported theoretical expressions for
electron occupancies \cite{notreprb}.  When $\epsilon \ll k_{B} T \ll
B \epsilon $ the Fermi-Dirac (FD) distribution is recovered with great
accuracy.  But $k_{B} T$ needs not be very large compared with
$\epsilon$ in mesoscopic devices, \emph{e.g.}, short quantum wires or
quantum dots.

Note that laser noise depends in general, not only on active levels
\emph{average} occupancies, but also on the fact that, even in the
equilibrium (or quasi-equilibrium) state, electrons keep moving in and
out these levels, causing the optical gain to fluctuate.  It is only
in the linearized theory (see the appendix of the present paper) that
statistical gain fluctuations may be ignored.  Such fluctuations are
automatically taken into account in Monte-Carlo simulations.

Let us now consider the process of thermalization between the electron
gas and the lattice.  To enforce thermalization each electron is
ascribed a probability $p$ per unit time of being demoted to the
adjacent lower level provided this level is empty, and a probability
$p q$, where $q \equiv \exp(-\epsilon / k_{B} T)$ of being promoted to
the adjacent upper level if it is empty.  Strictly speaking, this
thermalization model would be applicable to solids with $\hbar
\omega_{phonon} = \epsilon$, but the detailed modeling turns out to be
rather unimportant.\footnote{A more accurate modeling would require
that the optical phonons be pictured in their non-equilibrium state,
and be coupled to pairs of acoustical phonons.  The latter may
reasonably be expected to be at thermal equilibrium with the heat
sink.  It is clearly not the purpose of the present paper to go that
far.  A preliminary modeling suffices to explain the main concepts and
results.} If $p$ is large, thermalization is very efficient.  This
implies that electron-gas temperatures in both bands are equal to the
lattice temperature, i.e., are constant in the course of time.  The
main purpose of this paper is to consider the noise spectrum when $p$
is not large, in which case electron gas temperatures are ill defined. 
Near equilibrium, it is immaterial whether Auger or thermalization
transitions are dominant since both lead to well-defined temperatures
and, in the appropriate limit, to the FD distribution.  But because
lasers are out of equilibrium systems, intensity-noise spectra do
depend on which one of the Auger or thermalization processes
dominates.

Spontaneous radiative electronic decay (involving radiation into free
spa\-ce) may be accounted for by ascribing to each electron in the
conduction band a probability of dropping to some empty level in the
valence band.  The noise associated with this process is automatically
expressed by the Monte Carlo simulation.  For the sake of brevity,
this process is presently ignored.  Spontaneous decay may be neglected
when the driving current exceeds approximately ten times the threshold
current.

Consider now an isolated system consisting of a single-mode optical
cavity resonating at angular frequency $\omega$ with $\hbar
\omega = d \epsilon$, $d$ being an integer, and containing some electron
gas.  Precisely, we suppose that the coherent interaction takes place
between the middle of the conduction band and the middle of the
valence band, that is: $\hbar \omega = (G_{p} + B) \epsilon$.  A
prerequisite of the Monte Carlo method is that the optical field
enters only through the number $m$ of light quanta.  That is, photonic
superposition states, as well as electronic superposition states, are
not considered in the present theory.  Quantum Optics effects such as
trapped states, resonance fluorescence, and collective effects such as
superfluorescence are indeed ignored.  This conforms with the
classical rate equation treatment of lasers, found, for example, in
\cite{loudon, meystre}.  Recent rigorous calculations relating to the
mesomaser validate rate-equation methods by showing that even with few
atoms (or electron-hole pairs) typical Quantum Optics effects get
washed out \cite{elk}.  A particle-like Monte Carlo method applied to
optical amplifiers \cite{giulani} also shows convergence toward the
rate equation model.

Stimulated absorption is modelled by assigning a probability $m$ to
electrons in the lower working level to be promoted to the upper
working level (if that level is empty).  Stimulated emission is
modeled by assigning a probability $m + 1$ to electrons in the upper
working level to be demoted to the lower working level (if that level
is empty).  In the laser theory the ``1'' of the Einstein expression
$m + 1$ may be neglected in the steady-state because $m$ is a large
number.  The term ``1'' should be kept however in the Monte Carlo
simulation because the initial value of $m$ considered is 0.  Without
that term, laser start-up would not occur.  Setting as unity the
factor that multiplies the expressions $m$ or $m + 1$ amounts to
selecting a time scale.

Optical pumping (from a thermal source, for example) would be modeled
by assigning some constant probability to electrons in low
valence-band levels to be promoted to high conduction-band levels,
provided these levels are empty, \emph{and} almost the same
probability for the opposite transition.  In that case the pumping
rate fluctuations would be close to the shot-noise level.

But the electrical current generated by cold high-impedance electrical
sources is almost non-fluctuating as a consequence of the Nyquist
theorem, as was first shown by Yamamoto.  The nature of the detected
light depends on the ratio of this impedance to the intrinsic dynamic
resistance of the laser \cite{richardson, yamhaus}.  We restrict
ourselves to perfectly regular electrical pumping, \emph{i.e.} to
infinite cold impedances.  Quiet electrical pumping is modeled by
promoting low-lying electrons into high-lying levels
\emph{periodically} in time\footnote{If the lowest level happens to be
unoccupied or if the highest level happens to be occupied, a rather
infrequent circumstance, the program searches for the next adjacent
levels.}, every $\Delta t = 0.2$~ns, corresponding to a pumping rate
$J = 5 \text{ns}^{-1}$.  Because the time period considered is very
short in comparison with the time scales of interest, this
prescription implies that the pumping rate is nearly constant.  This
has been verified numerically.

Light quanta absorption is supposed to be due to the detector alone,
that is, no additional optical loss is being considered.  Without loss
of generality, the detector is supposed to be located inside the
optical cavity, as in many early classical laser-noise theories.  It
is in fact immaterial whether the detector is located inside the
resonator or is coupled to the cavity through some partially
transmitting mirrors, as long as no spurious reflection occurs. 
Detecting atoms are assigned a probability $\alpha m$ of being
promoted to the upper state, where $\alpha$ denotes a constant.  Once
in the excited state, detecting atoms are presumed to decay
non-radiatively back to their ground state.

\section{Laser Noise From a Birth-Death Process}
\label{birth-death}

The method is best explained by considering first time intervals
$\delta t$ small enough that the probability that an event of a
particular kind occurs within it is small compared with unity, and
that the probability of two or more events occurring is negligible,
\emph{e.g.}, $\delta t = 10^{-15}$\,s.  A typical run lasts
$T_{m} = 1\;\mu$s, corresponding to $10^9$ elementary time intervals. 
The total number of events per run is on the order of $10^8$. 
Averaging is made over 20 independent runs.  Instead of the above 
pedestrian approach, the algorithm actually employed accounts more 
rigorously for the birth-death process and minimizes the CPU time.

Stimulated decay of an electron during an elementary time interval
$\delta t$ is allowed to occur with probability $(m + 1) \delta t$, $m$
being incremented by $1$ if the event does occur.  Likewise,
stimulated electron promotion is allowed to occur with probability
$m \delta t$, $m$ being reduced by 1 if the event occurs.  The fact
that the proportionality constant has been omitted amounts to
selecting a time unit, typically, 1 ns.

Thermalization is required for steady-state laser operation.  In the
computer model, thermalization is ensured by ascribing to each
electron a probability $p \delta t$ to decay to the adjacent lower
level if that level is empty, and a probability $q p \delta t$ to be
promoted to the adjacent upper level if that level is empty, where $q
\equiv \exp(-\epsilon / k_{B} T)$ denotes the Boltzmann factor.  We
select $T = 100$~K, corresponding to $q = 0.891$.  Without absorption and
pumping ($\alpha = 0$, $J = 0$), the program gives level occupancies very
close to those predicted by the Fermi-Dirac (FD) distribution.

Only regular pumping is considered with an electron at the bottom of
the VB promoted to the top of the CB every $\Delta t = 0.2$~ns.  This
period corresponds to a pumping rate $J = 1/\Delta t = 5$ events per
ns, and a pump electrical current in the nA range.  For a
1~$\mu$m--long quantum wire with a 10~nm $\times$ 10~nm cross-section
this corresponds to $10^7$~A/cm$^{3}$.

Each light quantum is ascribed a constant probability $\alpha \delta
t$ of being absorbed by the photodetector, with $\alpha = 0.5 \;
\text{ns}^{-1}$.  The average number of light quanta in the cavity
follows from the average-rate balancing condition $J = \alpha
\overline{m}$, that is, $\overline{m} = 10$ light quanta.

The Fig.~\ref{fig2} illustrates three of these elementary processes by
means of a sequence of four frames extracted from a computer
simulation involving only 10 levels in each bands.  At the start, the
system has already reached a stationary regime.  A sample of the
electron distribution is shown on the left.  The corresponding time
and the number of light quanta stored in the cavity are respectively
$\tau_{0} = 0.9486$~ns and $m = 2$.  The first event at $\tau_{1} =
0.9488$~ns is a VB thermalization.  Its effect is to decrement the
system energy of $\epsilon$ since an electron is demoted by one energy
step.  The second event at $\tau_{2} = 1$~ns is an electrical pumping
event that promotes the VB electron occupying the lowest energy level
to the highest energy level of the CB. The third event illustrates
stimulated emission between the prescribed lasing levels at $\tau_{3}
= 1.0111$~ns.  As a result, the number of light quanta is incremented
from $m = 2$ to $m = 3$.

Except for the pumping events all the processes are governed by a
Poisson probability law.  It follows that the whole system aside from
pumping events also obeys a Poisson probability law.  Knowledge of the
laser microscopic state at time $\tau_{i}$ implies knowledge of the
next-jump density function.  An exact Monte Carlo simulation
\cite{gillespie} of the laser evolution is then easily obtained by
randomly picking up the next event time $\tau_{i+1}$ from a Poisson
law and, next, picking the event type from a uniform law weighted by
the count of potential events for each type.  This method is more
rigorous and more efficient than simulations based on infinitesimal
time steps.  The time required to obtain a photo-detection spectrum is
on the order of a few hours on desk computers.

The times $t_{k}$ of occurrence of photo-detection events are
registered once a steady-state regime has been reached, as is always
the case for the kind of lasers considered.  The detection rate $Q(t)$
is the sum over $k$ of $\delta(t-t_{k})$, where $\delta(.)$ denotes
the Dirac distribution.  Considering that the photo-detection events
are part of a stationary process, the two-sided spectral density of
the detection rate fluctuation $\Delta Q(t) \equiv Q(t) -
\overline{Q}$ is \cite{pap}
\begin{equation}
	\label{sq}
	S_{\Delta Q}(\Omega)=\frac{1}{T_{m}} \overline{\left|
	\sum_{k}\exp(-j\Omega t_{k}) \right|^{2}}
\end{equation}
where the overline stands for averaging and $\Omega \equiv 2 \pi
n/T_{m}$ with $n$ a nonzero integer.  Note that $2 S_{\Delta
Q}(\Omega) df$, where $f \equiv \Omega / 2 \pi$, represents the power
flowing out of a filter of width $df$ following the detector.  For
uniformly-distributed independent events, \emph{i.e.} for a Poisson
process, Eq.~(\ref{sq}) gives the shot-noise formula $S_{\Delta
Q}(\Omega) = \overline{Q}$.

\section{Numerical Results}
\label{results}

Simulations will be reported for three values of the thermalization
parameter, namely $p = 25\,000 \; \text{ns}^{-1}$, $1\,000 \;
\text{ns}^{-1}$, $250 \; \text{ns}^{-1}$, and $T=100$~K. Large values
of $p$ enforce well-defined, constant temperatures to the electron
gas.  Conversely, small values of $p$ correspond to large
carrier-lattice thermal resistances.

The total numbers of events during a run lasting 1\,$\mu$s are,
respectively, about $6.2\,10^8$, $3.9\,10^7$ and $1.1\,10^7$.  The
number of events of various kinds are listed in Table \ref{evts} for
the three values of $p$ considered.  The number of VB (resp., CB)
``cooling'' events is the number of downward electron transitions in
the valence band (resp., conduction band) due to thermalization. 
Likewise, VB (resp., CB) ``heating'' refers to upward electron shifts.

The following observations can be made: 

\begin{itemize}

\item The first two lines of the table show that, over a run, the
number of detection events is essentially equal to the number of
pumping events.  Since the number of pumping events does not vary from
run to run, it follows that the number of photo-detection events does
not vary either, \emph{i.e.}, the light is ``quiet''.  Non-zero
variances of the photo-count would appear only over much shorter
durations.

\item The next two lines show that the difference between the numbers
of stimulated emission and absorption events is nearly equal to the
number of photo-detection events.  Because of the band symmetry, the
numbers of stimulated events are almost independent of $p$.

\item The difference between the number of cooling and heating events
corresponds to the power delivered by the pump in excess of the power
removed by the detector.  This difference is almost independent of
$p$.

\end{itemize}

The CB level-occupancies are represented on the left-hand part of
Fig.~\ref{fig3}.  The CB electron-occupancies and the VB \emph{hole}
occupancies are symmetrical with respect to the middle of the bandgap. 

\begin{itemize}
    
    \item For $p = 25\,000 \; \text{ns}^{-1}$, electron occupancies
    are very close to the Fermi-Dirac (FD) distribution, except near
    the edges of the band.  A least-square fit shows that the carrier
    temperature is $T_{c} = 105.6$~K for both bands.  The quasi-Fermi
    levels (referred to the bottom of the bands) are respectively
    $\mu_{CB} = 60.2$ and $\mu_{VB} = 40.8$.

    \item For $p = 1\,000 \; \text{ns}^{-1}$, a fit gives $T_{c} =
    132$~K for both bands, $\mu_{CB} = 63.5$ and $\mu_{VB} = 37.5$. 
    There is a dip due to `spectral hole burning' (SHB) at the lasing
    level shown by an arrow.  This dip is difficult to see on the
    figure, but it nevertheless influences importantly the noise
    properties of the laser.

    \item For $p = 250 \; \text{ns}^{-1}$, the dip at the lasing level
    is conspicuous.

\end{itemize}

The right part of Fig.~\ref{fig3} gives a comparison between spectra
calculated from Monte Carlo data using Eq.~(\ref{sq}) and the spectrum
obtained from the elementary laser-diode noise theory (see the
appendix) using the same set of parameters, $(\alpha, T, \epsilon,
J )$.  The top part of the figure corresponds to efficient
thermalization.  The spectral density is below the shot-noise level up
to a frequency of 42~MHz.  Notice the strong relaxation oscillation. 
There is good agreement between the Monte Carlo simulation and the
linearized theory.

For moderate thermalizations the spectral density is below the
shot-noise level up to a frequency of 25~MHz and no more agrees with
the linearized theory.  An increase of temperature from 100~K to
132\,K does not suffice to reproduce the observed shift.  The change
in spectral density may be attributed to spectral-hole burning and
carrier heating.

The bottom curve corresponds to poor thermalization.  The relaxation
oscillation is strongly damped.  The frequency range where the
spectral density is below the shot-noise level now extends only up to
8~MHz.

\section{Conclusion}
\label{conclusion}

A Monte Carlo computer program keeping track of the occupancy of each
level in the conduction and valence bands and of the number of light
quanta in the optical cavity has been built.  It has been applied to
regularly-pumped mesoscopic laser diodes, having equally-spaced
one-electron levels in each band.  Numerical results have focused on
laser noise, especially the spectral density of the photo-detection
rate.  When the electron-lattice thermal contact is good, theoretical
results based on linearization and the Fermi-Dirac distribution are
recovered.  In particular, it is verified that sub-Poissonian light
may be obtained.  But when the thermal contact is poor, as is the case
when lasers are driven to high powers, the simple theory is found to
be inaccurate.  Some of the changes observed may be accounted for by
temperature increase and gain compression (due to spectral-hole
burning).  But unexpected effects are also found.  In particular, an
increase of the spectral density at low frequencies is noted.  The
program has been augmented to account for the intraband Auger effect
that tends to ensure that the electron and hole temperatures are well
defined, but possibly fluctuate differently in the course of time.  It
is also easy to take into account spontaneous carrier recombination
and excess optical losses.

\section{Appendix: Laser Diode Intensity Noise for the Case of
Evenly-Spaced Electronic Levels.}

A simple explicit expression for the spectral density of the
photo-detection rate is presently derived.  The theory of laser diode
intensity noise for the case of non-fluctuating (or quiet) pumps was
first given by Yamamoto and others \cite {richardson} in 1986.  A
significantly simpler but strictly equivalent theory, based on rate
equations, was subsequently given by Arnaud and Est\'eban \cite {arn3}. 
The present appendix is based on the latter with some changes in the
way the results are presented.  Next, it will be shown that for
electronic levels with even spacing $\epsilon$, the photo-detection
spectrum depends on only two parameters, namely the dimensionless
light quanta absorption constant $\alpha$ and a normalized pump rate
$J^{\star} \equiv J \epsilon / k_{B} T$.  As in the main text, the
parameter $J$ denotes the pumping rate equal within our approximations
to the average light-quanta output rate $Q$.  In the present appendix
ideal thermalization of the electron gas is assumed.

The main mechanisms involved in laser operation are stimulated
emission and stimulated absorption.  Einstein has shown that the
probability that an electron in the lower working level be promoted to
the empty upper working level is equal to the number $m$ of light
quanta (or photons) in the optical cavity, defined as the ratio of the
optical field energy and $\hbar \omega $, where $\hbar$ denotes the
Planck constant divided by $2 \pi$, and $\omega$ the angular frequency
of oscillation.  The probability that an electron in the upper level
be demoted to the empty lower level is equal to $m+1$.  These
relations hold to within a constant common factor that we set equal to
unity.  This amounts to selecting a time unit.  Because only far-above
threshold operation is considered we have: $m \gg 1$ so that the ``1''
in the expression $m+1$ may be neglected.  Spontaneous carrier
recombination is first neglected, but expressions for spontaneous
decay rates are given at the end of this appendix.  From the present
viewpoint, randomness in laser operation enters mainly because
stimulated transitions obey \emph{probability} rather than
deterministic laws.  The alternative interpretation of laser noise as
resulting from the field spontaneously emitted into the oscillating
mode, though plausible on some respects, does not seem able to lead in
a natural manner to sub-Poissonian output light statistics, see \cite
{arn3}.

In semiconductors it may happen that the two working levels are both
occupied or that both are empty, in which cases no transition may
occur.  Stimulated absorption takes place when the lower level is
occupied and the upper level is empty at an average rate denoted:
$A \equiv L(n) m$.  In the opposite situation, stimulated emission
occurs at an average rate $E \equiv G(n) m$.  Here $n$ denotes the total
number of electrons in the conduction band.  Because $m$ and $n$ are
large integers and their relative fluctuations are small above
threshold they may be viewed as continuous functions of time.  It is
also permissible to replace the average value of any smooth function
$f(m,n)$ by $f(\overline{m},\overline{n})$.  Overlines indicating that
average values are considered will be omitted when no confusion is
likely to arise, particularly in the expressions of spectral densities
and after linearization of the equations.  Stimulated absorption and
emission rates read respectively
\begin{equation}\label{rates}
  \mathcal {A}=L(n)m+a(t)\qquad \mathcal {E}=G(n)m+e(t)   
\end{equation}  
The independent white noise sources $a(t)$ and $e(t)$ express the
randomness of the transitions.  Their spectral densities are equal to
the average rates, that is, $S_{a} = A \equiv L(n) m$, $S_{e} =E \equiv
G(n) m$.
 
If the optical cavity-semiconductor system were isolated (no pump, no
optical absorption) the sum $n+m$ of the number of electrons $n$ in
the conduction band and of the number $m$ of light quanta in the
optical cavity would be a constant.  It would then suffice to know the
evolution equation for $m$ alone
\begin{equation}\label{m}
    \frac{dm}{dt}= \mathcal{R} \equiv \mathcal {E} - \mathcal {A}
    \equiv K(n) m + r(t)
\end{equation} 
where we have introduced the net gain
\begin{equation}\label{K}
    K(n) = G(n) - L(n) 
\end{equation} 
The average number $\overline{n}$ of electrons follows from the
steady-state condition: $K(\overline{n}) = 0$, while the average number
of light quanta $\overline{m}$ would depend on the energy initially
given to the isolated system.  Because absorption and emission
processes are independent, the spectral density of $r(t) \equiv
e(t) - a(t)$ is the sum of the average rates
\begin{equation}\label{noise}
    S_{r} = \left[ G(n) + L(n) \right] m
\end{equation} 

But lasers are in fact open systems with a source of energy (the pump)
and a sink of energy (the optical detector).  To obtain the evolution
equation for the number $m$ of light quanta one must subtract from the
right-hand-side of (\ref{m}) the loss rate $\mathcal{Q}$ due to the
detector, no other optical loss being presently considered.  Detection
is a linear process of average rate $Q = \alpha m$ involving a noise
term $q(t)$ whose spectral density equals the average rate. 
Accordingly
\begin{equation}\label {deltaq}
    \mathcal {Q}\equiv \alpha m+q(t)\qquad S_{q}=Q=\alpha m
\end{equation}
The parameter $\alpha$ represents the loss due to detector absorption. 
If the detector is located outside the cavity, $\alpha$ represents the
loss due to transmission through partially transmitting mirrors.

The equation describing the evolution of the number $n$ of electrons
in the conduction band, on the other hand, involves the constant pump
rate $J$ (quiet pump).  The two rate equations thus read
\begin{equation}\label{Laserm}
    \frac{dm}{dt}= \mathcal {R}- \mathcal{Q} 
\end{equation}        
\begin{equation}\label{Lasern}
    \frac{dn}{dt}= J-\mathcal {R} 
\end{equation}

In the steady-state, the right-hand-sides of (\ref{Laserm}) and 
(\ref{Lasern}) vanish, and therefore
\begin{equation}\label{j}
    J=K(\overline{n})\overline{m}=\alpha \overline{m}\equiv Q
\end{equation}
The above relation defines the steady-state values of $\overline{m}$
and $\overline{n}$, given $J = Q$, $\alpha$, and the $K(n)$ function.
    
For slow variations the left-hand-sides of the previous equations
(\ref{Laserm}) and (\ref{Lasern}) may be neglected and thus
$\mathcal{Q}(t) = J = \text{constant}$.  In other words, the detection
rate $\mathcal{Q}$ does not fluctuate when the pump is non-fluctuating
or ``quiet''.  This simple result holds because carrier losses and
light quanta losses, besides those due to detection, have been
neglected.  When the above equations are linearized and $\Delta m(t)
\equiv m(t) - \overline{m}$, $\Delta n(t) \equiv n(t) - \overline{n}$
are Fourier transformed, one obtains
\begin{equation}\label{Lm}
     j\Omega \Delta m=\gamma \Delta n +r-q
\end{equation} 
\begin{equation}\label{Ln}
     j\Omega \Delta n=-\gamma  \Delta n -K \Delta m  -r
\end{equation}
where $\Delta m$, $\Delta n$ are now functions of $\Omega$, and we have
introduced the net differential gain
\begin{equation}\label{ga}
    \gamma\equiv \frac {dK/dn}{K}
\end{equation}
Let us recall that $r$ and $q$ are uncorrelated white-noise processes. 
Their spectral densities, $S_{r}$ and $S_{q}$, are given in
(\ref{noise}) and (\ref{deltaq}), respectively.

The detection rate fluctuation $\Delta Q = \alpha \Delta m + q$ is
obtained from (\ref{Lm}) and (\ref{Ln}) after elimination of $\Delta
n$, as a weighted sum of the two independent white-noise terms $r$ and
$q$, in the form: $\Delta Q = a r + b q$, where $a$ and $b$ denote two
complex functions of $\Omega$.  The spectral density of $\Delta Q$ is
equal to $\lvert a \rvert^{2} S_{r} + \lvert b \rvert^{2} S_{q}$. 
Since the mathematical transformations has been given earlier
\cite{arn3} only the result is given below:
\begin{equation}\label{X}
    \frac {S_{\Delta Q}(\Omega)}{Q}=1+\frac{2n_{p}\Omega ^2-\gamma^2
    m^2 \alpha ^2 }{\gamma^2 m^2\Omega^2+(\gamma m\alpha -\Omega
    ^2/\alpha )^2}
\end{equation}
where
\begin{equation}\label{np}
    n_{p}=\frac{G(n)}{K(n)}   
\end{equation}  
denotes the population inversion (or ``spontaneous emission'') factor, 
and $Q$ the average detection rate. This expression vanishes at 
$\Omega = 0$ (quiet output). It is unity if $\Omega$ goes to infinity 
(i.e., the fluctuation is at the shot-noise level).

\begin{figure}[p]
	\centering
  	\includegraphics[scale=0.7]{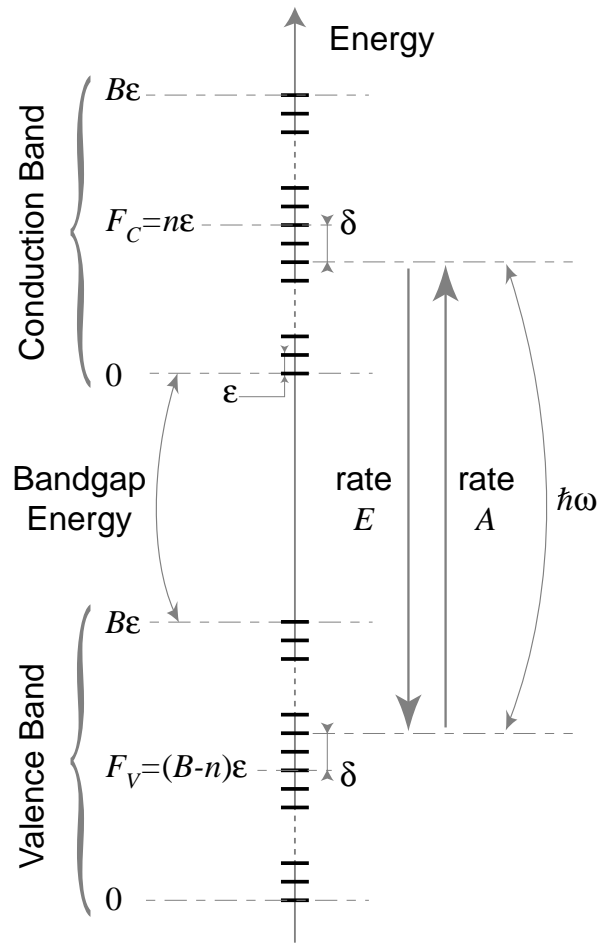}
	\caption{Schematic view of the energy level system
	considered.}
	\label{fig1}
\end{figure}

\begin{figure}[p]
	\centering
  	\includegraphics[scale=0.7]{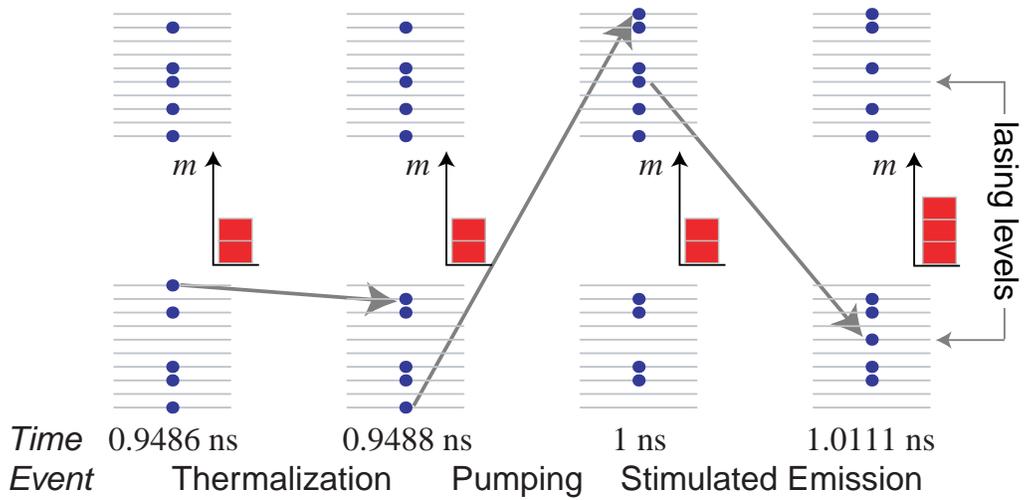}
	\caption{Four frames extracted from a Monte Carlo sequence for
	10 equally-spaced energy levels in VB and CB. Time in
	arbitrary units increases from left to right.  Insets give the
	number of light quanta $m$ stored in the cavity at some time. 
	Arrows indicate electron displacements from one energy level
	to another.}
	\label{fig2}
\end{figure}

\begin{figure}[p]
	\centering
  	\includegraphics[scale=0.7]{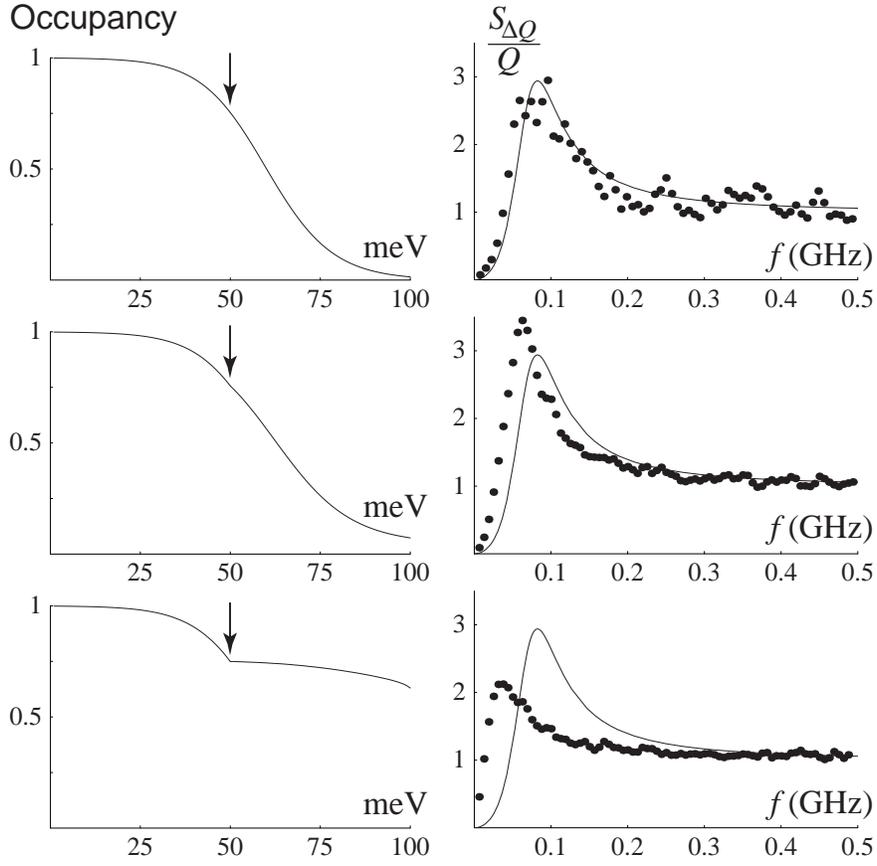}
	\caption{On the left, electron occupancy as a function of the
	energy referred to the bottom of the conduction band.  Arrows
	point to the lasing level.  On the right, normalized
	photo-detection spectra.  Dots are Monte Carlo results and
	solid lines are from (\ref{eq:fini}).  The thermalization
	rates are $p = 25\,000 \; \text{ns}^{-1}$ (top), $1\,000 \;
	\text{ns}^{-1}$ (middle), and $250 \; \text{ns}^{-1}$
	(bottom).}
	\label{fig3}
\end{figure}

Let us now evaluate the parameters $\gamma$ and $n_{p}$ entering into
the expression of the photo-detection rate spectrum.  As in the main
text model, each band is supposed to consist of $B$ one-electron
levels separated in energy by a constant $\epsilon$.  It is presently
assumed that the conduction and valence electron temperatures remain
at all times equal to the heat bath temperature $T$.  This amounts to
neglecting, besides temperature fluctuations, spectral-hole burning. 
The Fermi-Dirac distribution is applicable under the assumption that
$\epsilon \ll k_{B} T \ll B \epsilon$.  Under such conditions, the
total number $n$ of electrons in the conduction band may be evaluated
as if all levels above the quasi-Fermi level were empty and all levels
below it were fully occupied (and likewise for the valence band).  But
stable laser operation is possible only if the net gain depends
significantly on $n$.  It is therefore essential to employ the exact
Fermi-Dirac distribution when stimulated rates are being evaluated. 
Energies are counted upward from the bottom of the respective bands
(see Fig.~\ref{fig1}).  Because of charge neutrality we have
\begin{equation}\label{neps}
    n\epsilon =F_{c}=B\epsilon -F_{v}   
\end{equation}  
This relation determines the quasi Fermi levels as functions of the
number $n$ of electrons in the conduction band.

For definiteness, let us suppose that the two working levels are
located at the middle of the conduction and valence bands,
respectively, as shown on Fig.~\ref{fig1}, that is: $k_{c} = k_{v} = B/2$. 
It follows from (\ref{neps}) that
\begin{equation}\label{delta}
    \delta\equiv F_{c}-k_{c}\epsilon=k_{v}\epsilon-F_{v} 
\end{equation}
If we set
\begin{equation}\label{x}
    x\equiv \exp(-\frac{\delta}{k_{B}T})=\exp(\frac{B\epsilon
    /2-n\epsilon }{k_{B}T})
\end{equation}
the upper and lower working level occupancies read respectively,
according to the Fermi-Dirac formula
\begin{equation}\label{nc}
    n_{c}=\frac{1}{1+x}\qquad n_{v}=\frac{x}{1+x}   
\end{equation}  
Thus the stimulated gain and loss constants are
\begin{equation}\label{G}
    G(n)=n_{c}(1-n_{v})=\frac{1}{(1+x)^2}\qquad
    L(n)=n_{v}(1-n_{c})=\frac{x^2}{(1+x)^2}
\end{equation}  
and
\begin{equation}\label{k}
    K(n)\equiv G(n)-L(n)=\frac{1-x}{1+x}
\end{equation}
The steady-state condition $K(\overline{n}) = \alpha$ gives the
steady-state value of $x$:
\begin{equation}\label{xx}
    \overline{x}=\frac{1-\alpha}{1+\alpha}  
\end{equation}  
The population inversion factor thus reads
\begin{equation}\label{npp}
    n_{p}\equiv
    \frac{G}{K}=\frac{1}{1-\overline{x}^2}=\frac{(1+\alpha)^2}{4\alpha
    }
\end{equation}  

The differential gain parameter is obtained after derivation of $K(n)$
with respect to $n$ and rearranging as
\begin{equation}\label{gam}
    \gamma
    =\frac{dK/dn}{K}=\frac{2}{x^2-1}\frac{dx}{dn}=\frac{\epsilon
    }{k_{B}T} \frac{2x}{1-x^2} =\frac{\epsilon }{k_{B}T}
    \frac{1-\alpha^2}{2\alpha }
\end{equation}  
When the above expressions of $n_{p}$ and $\gamma$ are introduced in
(\ref{X}) we finally obtain the photo-detection rate spectral density
as a function of the frequency $f \equiv \Omega / 2 \pi$ with
parameters $\alpha$ and $J^{\star} \equiv J \epsilon / k_{B} T$
\begin{equation}
    \frac{S_{\Delta Q} ( f ) }{Q} = 1 +
    \frac{\frac{\alpha ( 1 + \alpha 
    )}{1-\alpha}\frac{F}{J^{\star}}-1}{\frac{1-\alpha^2}{2 
    \alpha^2}J^{\star} F + \left( 1 - F \right)^2}
    \label{eq:fini}
\end{equation}
where $F \equiv \left( f / f_{r} \right)^2$, $\left( 2 \pi f_{r} 
\right)^2 \equiv \frac{1-\alpha^2}{2}J^{\star}$ and $J^{\star} \equiv 
\frac{J \epsilon}{k_{B} T}$.

In the main text, very good agreement between the numerical simulation
for a large thermalization parameter and the above analytical formulas
was found.  Typical parameter values are $\alpha = 0.5$, $J^{\star}
\approx 0.58$.  Thus the relaxation frequency $f_{r} \approx
0.074$~GHz.

Within the present linearized theory, the spectrum would not be
affected if there were many upper working levels in the conduction
band instead of a single one, and likewise many lower working levels
in the valence band.  But when thermalization is poor the number of
working levels in each band needs to be specified.  Spectral-hole
burning is indeed of lesser importance when the number of working
levels is large.

If we suppose that spontaneous emission events satisfy the electron
momentum conservation law, spontaneous carrier recombination occurs at
an average rate $S(n) = \sigma n$ where the constant $\sigma$
depends on the optical density of mode, and thus on the dimensionality
of the active region.  This average rate should be supplemented with a
noise term $s(t)$ whose spectral density is equal to the average rate. 
Auger spontaneous carrier recombinations would obey different laws not
considered in this paper.  Note that Monte Carlo simulations
automatically account for all the noise sources associated with
transition events.

\section*{Acknowledgments}

This work was supported by the STISS Department of Universit\'e
Montpellier II. The authors acknowledge helpful discussions with F.
Philippe and J.-M. Bo\'e.

\begin{table}[p]
\caption{Number of events of various kinds during a $1\;\mu$s-run}
\begin{center}
\begin{tabular}{ r r r r }
\hline
$p$ (ns$^{-1}$)		& 25\,000 	& 1\,000 	& 250\\
\hline\hline
Pumping 		& 5\,000	& 5\,000	& 5\,000 \\
Detection		& 5\,004	& 5\,000	& 5\,000 \\
Stimulated abs.		& 882		& 865		& 876 \\
Stimulated emi.		& 5\,883	& 5\,865	& 5\,876 \\
VB ``cooling''		& 155\,286\,151	& 9\,819\,009	& 2\,848\,060\\
VB ``heating''		& 155\,036\,238	& 9\,569\,413	& 2\,619\,666\\
CB ``cooling''		& 155\,303\,449	& 9\,779\,404	& 2\,824\,115\\
CB ``heating''		& 155\,058\,595	& 9\,534\,806	& 2\,599\,576\\
\hline
\end{tabular}
\end{center}
\label{evts}
\end{table}

\clearpage

\end{document}